\documentclass[twocolumn,showpacs,preprintnumbers,amsmath,amssymb]{revtex4-1}

\usepackage{graphicx}
\usepackage{dcolumn}
\usepackage{bm}
\usepackage{color}
\usepackage{float}

\begin{document}

\preprint{APS/123-QED}

\title{Helimagnetic Structure and Heavy-Fermion-Like Behavior \\
in the Vicinity of the Quantum Critical Point in Mn$_3$P}

\author{Hisashi Kotegawa$^{1}$, Masaaki Matsuda$^2$, Feng Ye$^2$, Yuki Tani$^1$, 
Kohei Uda$^{1}$, \\
Yoshiki Kuwata$^{1}$, Hideki Tou$^1$, Eiichi Matsuoka$^1$, Hitoshi Sugawara$^1$, 
Takahiro Sakurai$^3$, Hitoshi Ohta$^{1,4}$, Hisatomo Harima$^1$, Keiki Takeda$^5$, Junichi Hayashi$^5$, Shingo Araki$^6$, and Tatsuo C. Kobayashi$^6$}

\affiliation{
$^{1}$Department of Physics, Kobe University, Kobe 658-8530, Japan \\
$^{2}$Neutron Scattering Division, Oak Ridge National Laboratory, Oak Ridge, TN 37831\\
$^{3}$Research Facility Center for Science and Technology, Kobe University, Kobe, Hyogo 657-8501, Japan \\
$^{4}$Molecular Photoscience Research Center, Kobe University, Kobe, Hyogo 657-8501, Japan \\
$^{5}$Muroran Institute of Technology, Muroran, Hokkaido 050-8585, Japan \\
$^{6}$Department of Physics, Okayama University, Okayama 700-8530, Japan
}

\date{\today}

\begin{abstract}

Antiferromagnet Mn$_3$P with Neel temperature $T_N=30$ K is composed of Mn-tetrahedrons and zigzag chains formed by three inequivalent Mn sites.
Due to the nearly frustrated lattice with many short Mn-Mn bonds, competition of the exchange interactions is expected.
We here investigate the magnetic structure and physical properties including pressure effect in single crystals of this material, and reveal a complex yet well-ordered helimagnetic structure.
The itinerant character of this materials is strong, and the ordered state with small magnetic moments is easily suppressed under pressure, exhibiting a quantum critical point at $\sim1.6$ GPa.
The remarkable mass renormalization, even in the ordered state, and an incoherent-coherent crossover in the low-temperature region, characterize an unusual electronic state in Mn$_3$P, which is most likely effected by the underlying frustration effect. 

\end{abstract}

\maketitle

The properties of interacting conduction electrons deviate from those of free electrons.
Such interactions often result in renormalization of the electron mass, which is the essence in strongly correlated electron systems.
They are generally remarkable near an instability, such as itinerant-localized crossover like the Kondo effect, the proximity of a Mott insulator, and criticality of some degrees of freedom.
Rich physics has been developed in such backgrounds including their interplay, and quantum criticality has been enthusiastically investigated for its ability to induce peculiar behaviors of electrons \cite{Lohneysen,Gegenwart,Shibauchi,Furukawa}.


Magnetic frustration is also an important factor to induce a wide variety of physical phenomena.
It can arise from geometrical constraint, or competition of various short-range exchange interactions. 
The effect is generally not well-established in itinerant magnetic systems, because it is thought to be weakened by long-range interactions. 
In itinerant systems, an aspect brought by competition of exchange interactions is a stabilization of helical magnetic structure, as has long been discussed in systems such as MnP-type materials \cite{Takeuchi,Kallel,Dobrzynski,Yano}. 
Additionally, in such systems, the Dzyaloshinsky-–Moriya (DM) interaction is often effective.
It is also short-range exchange interaction, and contribution to the detailed helical structure has been suggested in MnP \cite{Yamazaki}. 
The competing exchange interactions and the DM interaction are characterized by crystal symmetry, general to many materials, and are the key inducers of rich and intriguing phenomena \cite{Shiomi,Nakatsuji_Mn3Sn,Kurumaji}.
However, their role in the itinerant regime is poorly understood and remains an intriguing issue. 
Particularly, an interplay of the frustration and quantum phase transition has not been sufficiently explored in $d$-electron systems.

\begin{figure}[htb]
\centering
\includegraphics[width=0.95\linewidth]{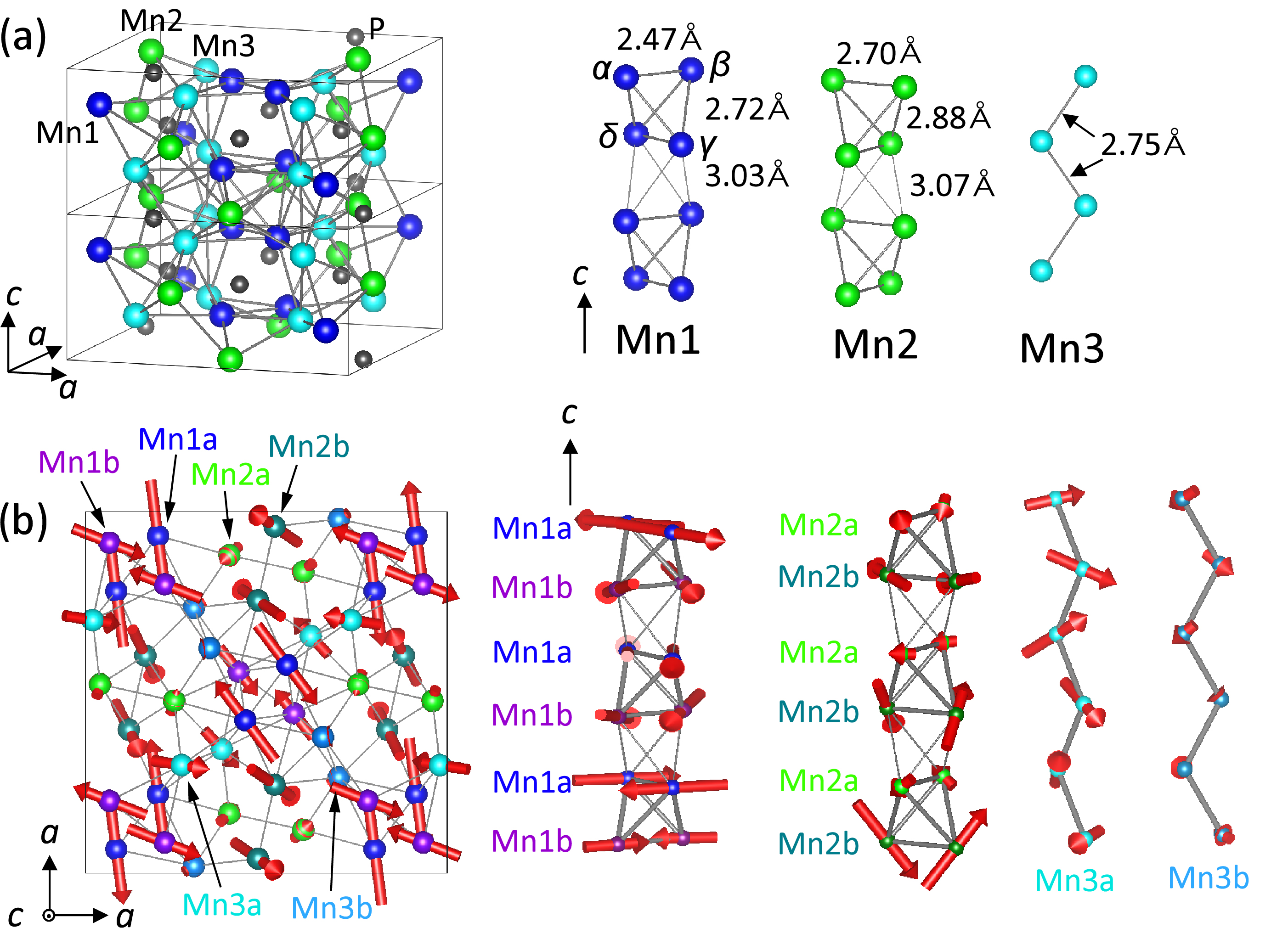}
\caption[]{(color online) (a) Crystal structure of noncentrosymmetric Mn$_3$P. Two unit cells are shown. Tetragonal structure of space group $I$\=4 is composed of three Mn sites and one P site. The Mn1 and Mn2 sites form the tetrahedrons, while the Mn3 sites form the zigzag chain. (b) The helical magnetic structure of Mn$_3$P was determined at 7 K and at ambient pressure. Three Mn sites are separated into six sites with different modulations of their magnetic moments. Nearly AF couplings between the magnetic moments almost lying in the $ab$ plane are seen at the Mn1a and Mn1b sites, respectively. The $c$-axis components are significant at the Mn2 and Mn3 sites.}
\end{figure}

Expecting a remarkable frustration effect and an induced quantum phase transition, we spotlighted Mn$_3$P, whose properties have not been investigated sufficiently so far.
An earlier study of polycrystalline samples had suggested an antiferromagnetic (AF) transition at $T_N=115$ K \cite{Gambino}, but a recent report has corrected this result to $T_N = 30$ K \cite{Liu}.
In a latter paper, the M\"{o}ssbauer spectroscopy and neutron diffraction measurements on slightly Fe-substituted samples suggested an ordered moment below 1 $\mu_B$, indicating a remarkable itineracy of the Mn 3$d$ electrons.
However, the details of the magnetic structure are unknown \cite{Liu}.

Mn$_3$P crystallizes in a noncentrosymmetric tetragonal structure in the $I$\=4 space group (No. 82, $S_4^2$) \cite{Rundqvist}.
The crystal structure comprises three inequivalent Mn sites and one P site (see Fig.~1(a)).
The four Mn1 sites, denoted as $\alpha-\delta$ in Fig.~1, are equivalent and form a tetrahedron.
By the symmetrical operation of \=4, the ``$\alpha-\gamma$'' bond is equivalent to the ``$\gamma-\beta$'' bond, leading to an isosceles triangle composed of $\alpha, \beta$ and $\gamma$.
The Mn2 sites are similar in structure to the Mn1 sites, whereas the Mn3 sites form a zigzag chain.
Another important feature is the short distances between the inequivalent Mn sites: 2.70 \AA\ for Mn1-Mn2, 2.54 \AA\ for Mn1-Mn3, and 2.55 \AA\ for Mn2-Mn3.
Fifteen different types of Mn-Mn bonds exist within 3.0 \AA.
Competition of the exchange interactions is expected among so many Mn-Mn bonds.

In this single-crystal study, we revealed that the itinerant antiferromagnet Mn$_3$P possesses unusual physical properties, such as a complex helimagnetic structure accompanied by a double transition and pressure-tuned quantum criticality accompanied by heavy-fermion-like behavior.
The likely cause of these features is the frustration effect induced by the underlying structure.

Single crystals of Mn$_3$P were grown by the self-flux method, as described in the Supplemental Material \cite{SM}.
We performed a range of experiments including resistivity, susceptibility, specific heat, neutron scattering, and NMR measurements. The resistivity and neutron scattering measurements were also conducted under pressure. We also calculated the band structure of Mn$_3$P. 
The experimental method, NMR results and the band structure calculation are described in the Supplemental Material \cite{SM}.

\begin{figure}[htb]
\centering
\includegraphics[width=0.85\linewidth]{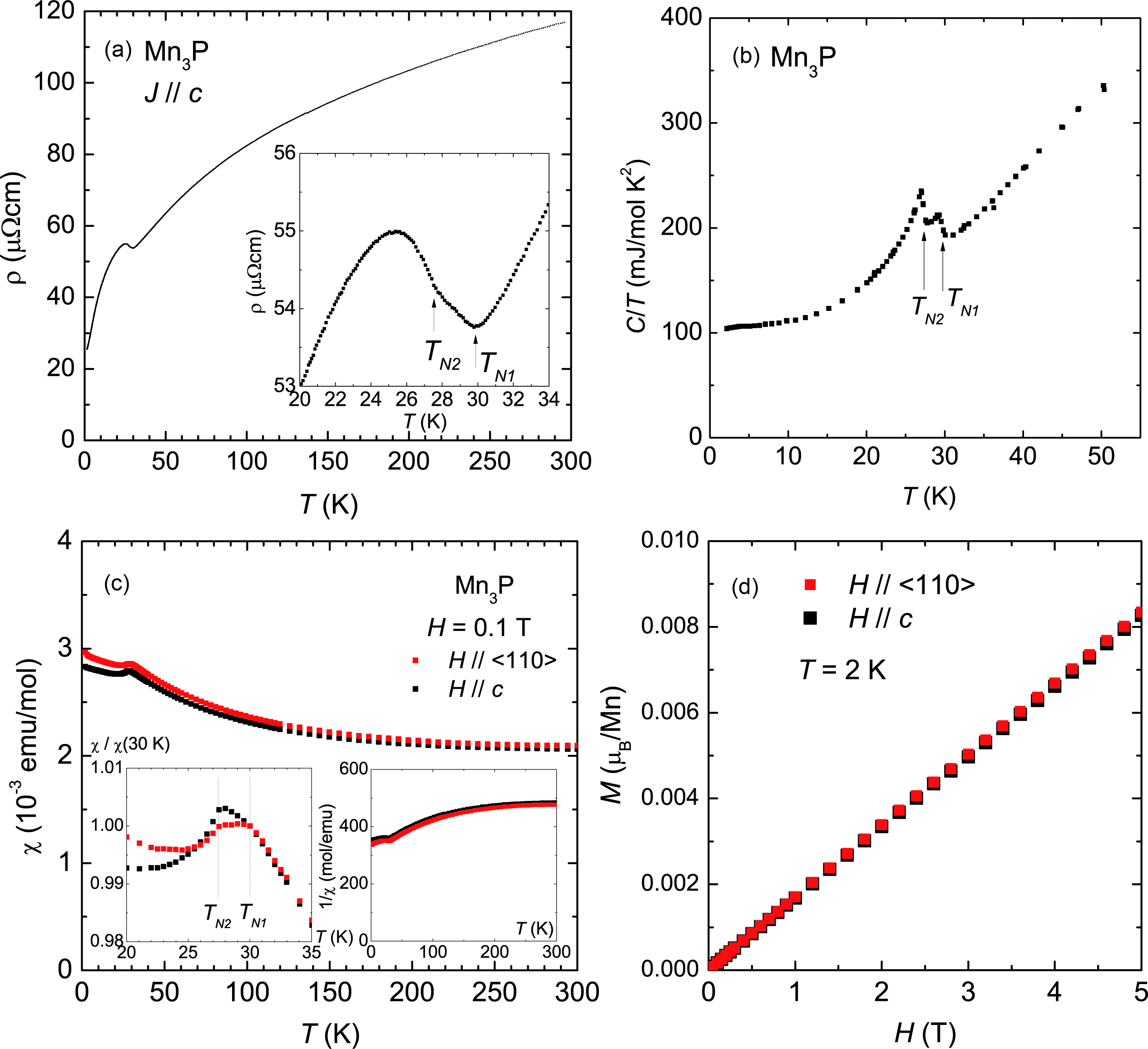}
\caption[]{(color online) Temperature dependence of (a) electrical resistivity, (b) $T$-divided specific heat ($C/T$), and (c) magnetic susceptibility at 0.1 T. All measurements detected two magnetic transitions; one at $T_{N1}=30$ K, the other at $T_{N2}=27.5$ K. The characteristic convex curve of $\rho$ above $T_{N1}$ is reminiscent of $f$-electron heavy fermion systems. (d) Magnetization curves at 2 K measured after field cooling. }
\end{figure}

Figures 2 (a-c) plot the temperature dependences of (a) electrical resistivity ($\rho$), (b) $T$-divided specific heat ($C/T$), and (c) magnetic susceptibility ($\chi$) of Mn$_3$P at ambient pressure.
The $\rho$ was characteristically convex below room temperature and exhibited two anomalies at $T_{N1}=30$ K and $T_{N2}=27.5$ K.
The $C/T$ demonstrated that both anomalies were second-order-like phase transitions.
They were also confirmed by successive suppressions in $\chi$.
As shown in the magnetization curve after field cooling to 2 K (Fig. 2(d)), no spontaneous magnetization occurred in the ground state.
Consistent with a previous report \cite{Liu}, the $\chi$ did not follow Curie-Weiss behavior (Fig.~2(c), inset), indicating the itinerant nature of the magnetism. 
The intrinsic nature of the double transition was further confirmed by NMR measurements \cite{SM}, and the neutron scattering data shown below.

\begin{figure}[b]
\centering
\includegraphics[width=0.9\linewidth]{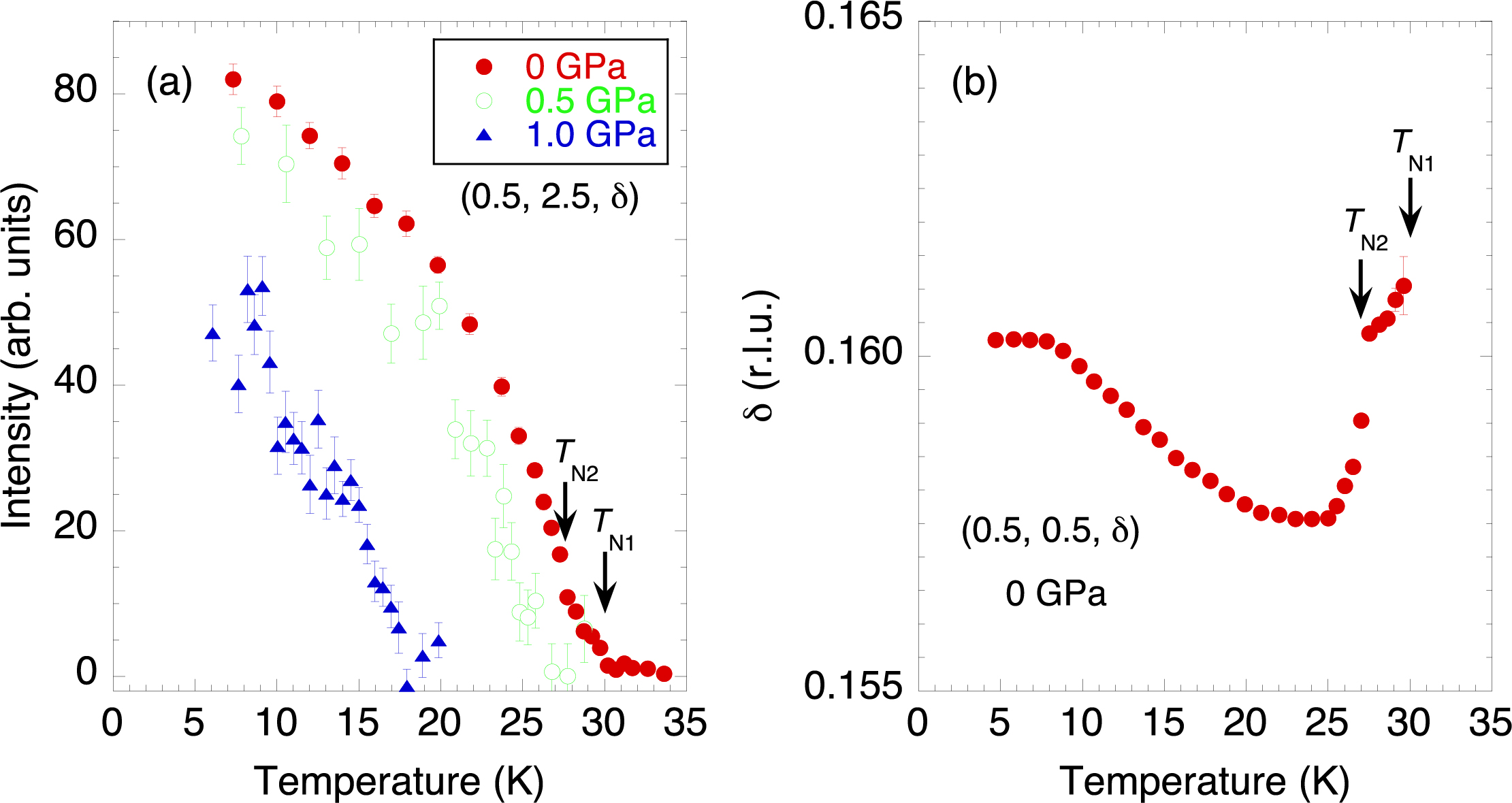}
\caption[]{(color online) (a) Temperature dependence of the magnetic Bragg intensities at (0.5, 2.5, $\delta$) with $\delta \sim0.16$, measured at different pressures. At ambient pressure, the magnetic scattering appears below $T_{N1}$ and exhibits a small kink at $T_{N2}$ (indicated by arrows). The two transitions are poorly resolved under pressure, but the scattering is observed at the same wave vector. (b) Temperature dependence of the incommensurability $\delta$ of the magnetic wave vector (0.5, 0.5, $\delta$) at ambient pressure.}
\end{figure}

\begin{figure}[b]
\centering
\includegraphics[width=\linewidth]{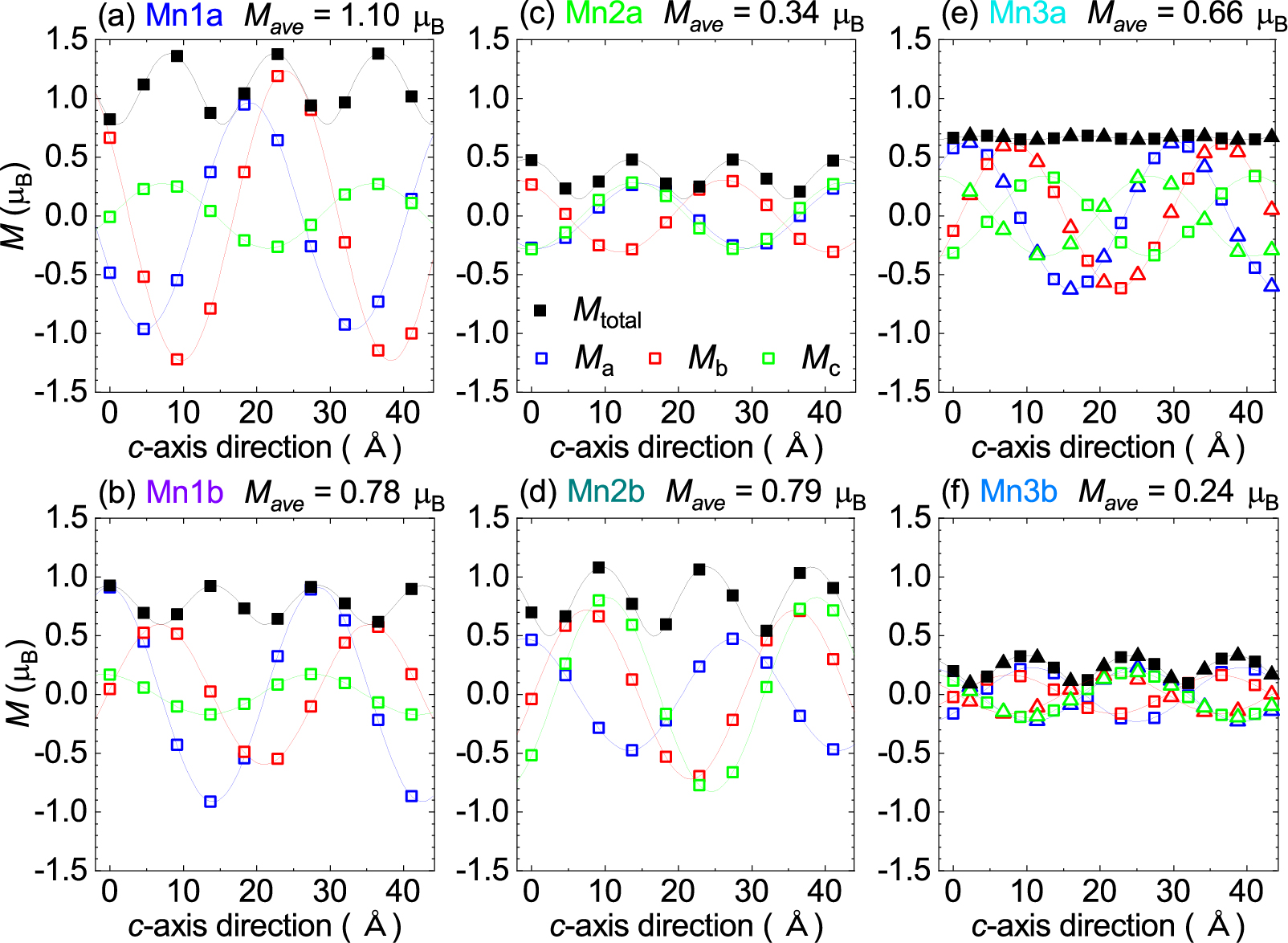}
\caption[]{(color online) Variation of the magnetic-moment components $M_a$, $M_b$, $M_c$, and $M_{total}$, along the $c$ axis at six Mn sites, measured at 7 K under ambient pressure. $M_{total}$ shows a clear ellipticity of the helix at most Mn sites. $M_{ave}$ denotes the average value at each Mn site. At the Mn1a and Mn1b sites, the magnetic moments almost lie in the $ab$ plane, but the moments are remarkably tilted at other sites. At the Mn3 sites, the squares and triangles indicate the two sublattices forming the zigzag chain. AF couplings between the sublattices appear in $M_c$ ($M_{a,b}$) at the Mn3a (Mn3b) sites. The Fullprof refinement was obtained as $R_F=11.3$ \%.}
\end{figure}

The magnetic structure of Mn$_3$P in the ground state was determined by neutron scattering measurements.
The magnetic wave vector was found to be ${\bf Q} = (0.5, 0.5, \delta)$ with $\delta \sim0.16$, which differs from the commensurate ${\bf Q}$ suggested in Ref.~14.
Fig.~3(a) shows the temperature dependence of the scattering intensity at ${\bf Q} = (0.5, 2.5, 0.16)$.
At ambient pressure, the intensity started increasing below $T_{N1}$ and exhibited a small kink at $T_{N2}$, similarly to our NMR results \cite{SM}.
The magnetic structure was refined using the 269 magnetic reflections measured at ambient pressure.
We utilized two programs: the magnetic structure shown in Fig.~1(b) as given by Jana \cite{Jana}, and the modulations of the magnetic moment along the $c$ axis given by Fullprof \cite{Fullprof} and shown in Fig.~4.
The resultant magnetic structures are consistent between the two programs. 
The crystal and magnetic-structure analyses are provided in the Supplemental Material \cite{SM}.
In the ordered state shown in Fig.~1(b), each Mn site is split into two Mn sites with independently modulated amplitudes and phases.
Mn1a and Mn1b (Mn2a and Mn2b) alternate along the $c$ axis, and Mn3a and Mn3b form different zigzag chains.
The complex non-collinear structure maintains a two-fold rotational symmetry $C_2$.
At most of the Mn sites, the size of the total magnetic moment oscillates along the $c$ axis (see Fig.~4), indicating clear ellipticity of the helix.
The origin of elliptical helices in $d$-electron systems is an interesting problem, and has been discussed in FeAs \cite{Rodriguez,Frawley}.
At all Mn sites, the size of the magnetic moment was about 1 $\mu_B$ or less, and was extremely small at the Mn2a and Mn3b sites.
At the Mn1a and Mn1b sites, the magnetic moments almost presented in the $ab$ plane, and were approximately coupled in antiparallel between the shortest ``$\alpha-\beta$'' (Mn1a) and the ``$\gamma-\delta$'' (Mn1b) bonds owing to $C_2$ symmetry, indicating a dominance by AF exchange interactions.
In the isosceles triangle composed of $\alpha, \beta$ and $\gamma$, the AF interaction in the ``$\alpha-\beta$'' bond frustrates the equivalent interactions in the ``$\beta-\gamma$'' and ``$\gamma-\alpha$'' bonds, similarly to geometrical frustration.
This yields a significant contribution to disturb the collinear magnetic ordering.
At the Mn2 and Mn3 sites, the magnetic moments have substantial $c$-axis components.
This helical state with the split sites, in which the sizes of the moments are significantly different, is conjectured to arise from reduced competition of the exchange interactions among the many Mn-Mn bonds. 
Contribution of the DM interaction is not clearly seen between the shortest Mn1-Mn1 bonds, which are almost coupled in antiparallel, but it will be important for reproducing the overall magnetic structure.
At $T_{N2}$, the $\delta$ abruptly but continuously changed by $\sim$2 \% \cite{SM}. 
The anisotropy of the averaged magnetic moment also changed when passing $T_{N2}$ \cite{SM}, but the weak magnetic intensities in the intermediate phase preclude a detailed analysis of the magnetic structure.
We consider that the six Mn sites undergo partial ordering at $T_{N1}$ and that the disordered Mn sites are ordered below $T_{N2}$.

The neutron scattering experiment clarified that all the Mn sites possess static magnetic moments, but the ordered states are unusual.
At low temperatures, the $C/T$ yielded a large electronic specific heat coefficient $\gamma = 104$ mJ/mol K$^2$, as shown in Fig.~2(b), corresponding to $\gamma_V = 3.6$ mJ/cm$^3$ K$^2$ per volume.
This is one of the larger values in $d$ electron systems \cite{Hussey}.
The $A$ coefficient of the $T^2$ term in the resistivity was also large value ($A=0.52$ $\mu\Omega$cm/K$^2$ at ambient pressure). 
The $A/\gamma^2$ ration was $4.8\times10^{-5}$, of the same order as the Kadowaki-Woods ratio.
In the band calculation in the PM state \cite{SM}, $\gamma_{\rm band}$ was estimated as 16.3 mJ/mol K$^2$ for a formula cell.
The experimentally obtained $\gamma$ in the ordered state was remarkably enhanced, suggesting that strong mass renormalization occurs in Mn$_3$P and survives even in the magnetically ordered state.

\begin{figure}[htb]
\centering
\includegraphics[width=\linewidth]{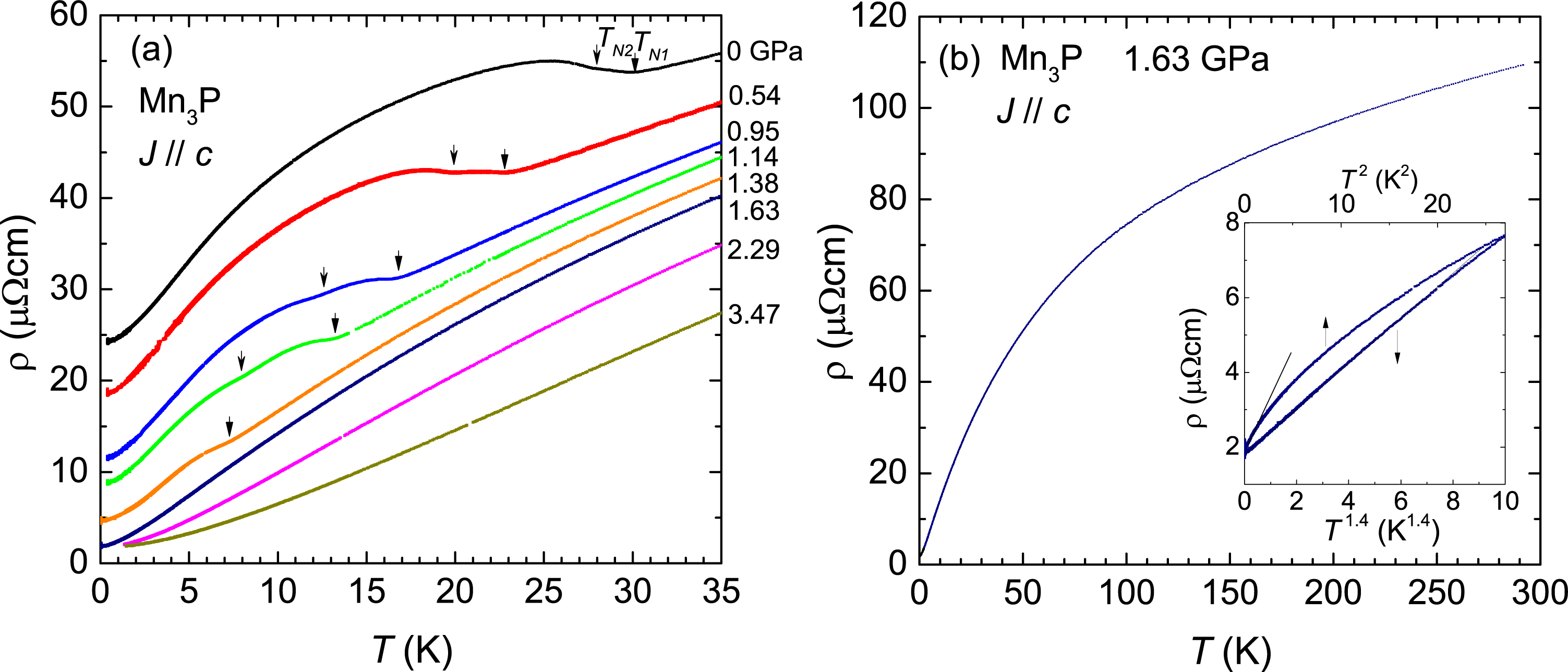}
\caption[]{(color online) (a) Electrical resistivity measured under pressure. The two transitions and the residual resistivity are simultaneously suppressed by applying pressure. (b) Resistivity behavior in the vicinity of the QCP. The residual resistivity ratio (RRR) is $\sim58$. The characteristic convexity appears over a wide temperature range. Inset: The resistivity at low temperature deviates from FL behavior, and obeys a $T^{1.4}$ dependence. The $\rho$ vs $T^2$ plots at several pressures are shown in the Supplemental Material \cite{SM}.}
\end{figure}

The pressure application drastically changed the electronic state of Mn$_3$P, as shown in Fig.~5(a).
Both transitions (indicated by arrows) were quickly suppressed under pressure.
The anomaly at $T_{N2}$ disappeared above $\sim1.4$ GPa, whereas $T_{N1}$ tended to zero at $1.5-1.6$ GPa.
The residual resistivity decreased drastically to $\sim2$ $\mu \Omega$cm at $\sim1.6$ GPa, and was thereafter independent of pressure.
Obviously, the large residual resistivity $\rho_0$ is inherent in the ordered state, although its origin is unclear.
The pressure-independent $\rho_0$ in the PM state probably originates from imperfectness in the sample, and its small value indicates a high-quality single crystal.
Figure 6 shows the pressure-temperature phase diagram of Mn$_3$P, and the pressure dependences of $\rho_0$ and $A$.
The continuous suppression of the ordered state suggests a quantum critical point (QCP) at $P_c \sim 1.6$ GPa.
Among Mn-based systems, Mn$_3$P is a rare example of an easily-inducible QCP; thus far, quantum phase transitions have been reported in only a few materials \cite{Takeda,Pfleiderer,Cheng_MnP}.
Another QCP, where $T_{N2}$ reaches 0 K, is also expected at $\sim1.4$ GPa. 
The phase diagram suggests that two ordered phases are almost degenerate at any pressure.
The pressure dependence of the magnetic Bragg intensities is shown in Fig.~3(a).
The incommensurate structure with $\delta$=0.16 was robust and no pressure dependence was observed up to 1 GPa; specifically, $\delta$ was 0.1608(2) at 0 GPa, 0.160(1) at 0.5 GPa, and 0.162(2) at 1 GPa.
The ordered magnetic moment gradually reduced with increasing pressure, reaching $\sim72$\%\ of its ambient-pressure value (on average) at 1 GPa.

\begin{figure}[htb]
\centering
\includegraphics[width=0.65\linewidth]{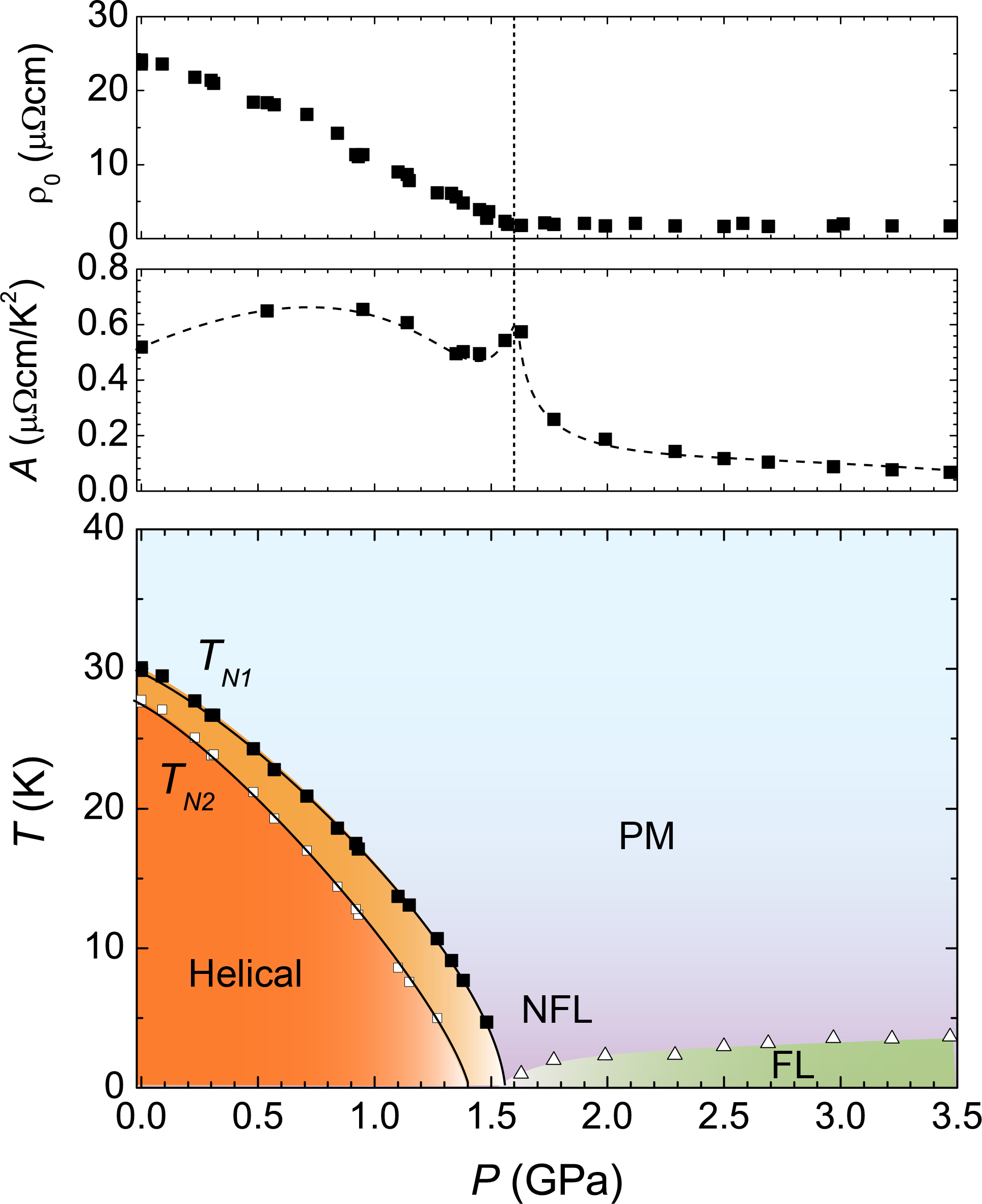}
\caption[]{(color online) Pressure dependences of $\rho_0$ and $A$ coefficient, and pressure-temperature phase diagram of Mn$_3$P. In the vicinity of QCP, the NFL behavior is dominant. The $A$ value is already large in the ordered state and shows a small peak at the QCP. }
\end{figure}

Figure 5(b) plots the temperature dependence of resistivity at 1.63 GPa, just above $P_c$.
The resistivity in the PM state was convex over a wide temperature range, reminiscent of $f$-electron heavy fermion systems.
The resistivity remarkably decreased below $50-100$ K, indicating a broad incoherent-coherent crossover in the low temperature region. 
As shown in the inset of Fig.~5(b), the resistivity at 1.63 GPa obeyed a $T^{1.4}$ dependence at low temperatures.
The vicinity of the QCP was dominated by NFL behavior and a distinct peak in the $A$, although $A$ was already large in the ordered state.
This indicates an unusually small entropy release through $T_{N1}$ and $T_{N2}$. 
The $A$ was suppressed at pressures above $P_c$, and the FL state became stabilized.
The observed $A$ in Mn$_3$P ($0.5-0.6 \mu \Omega$cm/K$^2$) is comparable to those of $f$-electron heavy fermion systems, and is $\sim30$ times larger than that of MnP, even near the QCP \cite{Cheng_MnP} where the helimagnetic phase terminates \cite{Matsuda_MnP,Wang_MnP}.

Several materials in $d$-electron systems have been reported as heavy fermion systems with strong mass renormalization.
Examples are $\beta$-Mn, (Y,Sc)Mn$_2$, LiV$_2$O$_4$, (Ca,Sr)RuO$_4$, Na$_{1.5}$Co$_2$O$_4$, CaCu$_3$Ir$_4$O$_{12}$, and $A$Fe$_2$As$_2$ ($A=$K, Cs) \cite{Shinkoda,Shiga,Kondo,Urano,Nakatsuji,Miyoshi,Cheng,Wu}. 
Such materials commonly exhibit a low characteristic temperature $T^*$, which generally gives a large $\gamma$ proportional to $1/T^*$.
Mn$_3$P should be included in this category.
In the above oxides and iron pnictides, $\chi$ obeys Curie-Weiss behavior at temperatures above $T^*$ \cite{Kondo,Nakatsuji,Miyoshi,Cheng,Hardy}, suggesting the presence of localized moments at high temperatures. 
Accordingly, $T^*$ is generally interpreted as a localized-itinerant crossover, broadly analogous to that observed in the $f$-electron systems.
The localized moment in $d$-electron systems primarily arises from proximity of a Mott insulator \cite{Arita,Jonsson,Anisimov,Hardy}.
However, in Mn$_3$P, the deviation of $\chi$ from Curie-Weiss law and the band calculation \cite{SM} suggest the obvious itinerant character of the $3d$ electrons.
It is unlikely that Mn$_3$P is compatible with models based on localized-itinerant crossover.

Heavy $d$-electrons may also arise from the effects of geometrical frustration, as discussed for $\beta$-Mn, Y$_{0.95}$Sc$_{0.05}$Mn$_2$ and LiV$_2$O$_4$.
They are divided into models with localized moments \cite{Fulde,Lacroix,Hopkinson,Burdin,Shannona,Laad} and itinerant models \cite{Pinettes,Tsunetsugu,Yamashita}. 
In the itinerant models, it has been proposed that frustration enhances the fluctuations of degenerate $t_{2g}$ orbitals \cite{Tsunetsugu,Yamashita}, but Mn$_3$P prohibits such degeneracy in principle, because the local symmetry of each Mn site is low as represented by its notation 1 ($C_1$).
Another route to heavy mass is degeneracy of the magnetic correlations inherent in frustration \cite{Pinettes}.
This route may be qualitatively adapted to Mn$_3$P, which is not a geometrically frustrated lattice but exhibits the unusual features of a strong frustration effect, such as the almost-degenerated magnetic transitions and the splitting into six Mn sites with quite different magnetic moment. 
In contrast to other itinerant heavy-fermion systems, $\beta$-Mn and Y$_{0.95}$Sc$_{0.05}$Mn$_2$, Mn$_3$P is a clean system, which excludes the possibility that the disorder effect enhances the incoherency.
The observed features in Mn$_3$P are worthy of further theoretical and experimental elucidations.

In conclusion, we performed comprehensive experimental characterizations and calculations of antiferromagnet Mn$_3$P.
We first identified a complex helimagnetic structure, in which three Mn sites are separated into six sites of different sizes and directions of their magnetic moments.
The helimagnetic structure accompanied by the double transition suggests competition of the exchange interactions, or frustration, in Mn$_3$P.
Second, the QCP was easily induced under pressure, a rare behavior in Mn-based materials.
Third, Mn$_3$P behaved similarly to $f$-electron-heavy fermion systems, namely, it presented a convex resistivity curve, a larger electronic specific heat coefficient than that estimated by the band-structure calculation, and a large $A$ coefficient of the resistivity
They demonstrate a strong mass renormalization in Mn$_3$P.
These noteworthy findings reveal that the itinerant system Mn$_3$P comprehends the areas of magnetism with competing interactions, a quantum criticality, and $d$-electron heavy fermion. 
It is an excellent example to merge them and to induce the novel interplay.

\section*{Acknowledgements}

We thank Kazuto Akiba for experimental supports.
This work was supported by JSPS KAKENHI Grant Number JP15H05882, JP15H05885, JP18H04320, and 18H04321 (J-Physics), 15H03689.and 15H05745.
This research used resources at the High Flux Isotope Reactor and the Spallation Neutron Source, DOE Office of Science User Facilities operated by the Oak Ridge National Laboratory.

\clearpage

\begin{center}
{\large - Supplemental Material-}

\end{center}

\section{Sample preparation}

Single crystals of Mn$_3$P were synthesized using a self-flux method.
A mixture of Mn : P = 4 : 1 was put into Al$_2$O$_3$ crucible and sealed in an evacuated quartz ampoule.
The ampoule was heated slowly up to 800 $^{\rm o}$C and held there for 4 days to promote a homogeneous reaction.
After that, the ampoule was heated up to 1100 $^{\rm o}$C for 2 hours and then cooled to 950 $^{\rm o}$C for 6 hours.
After centrifuging the flux at 980 $^{\rm o}$C, the bar-shape single crystals were obtained.
The crystal tends to grow along the $c$ axis and the typical size is $1 \times 1 \times 5$ mm$^3$. 
The crystals are robust against air and water, but sensitive to acid.
The remaining flux was removed mechanically from the crystal.

For the single-crystal x-ray diffraction measurement, the data were collected to a maximum $2\theta$ value of $\sim 61.4^{\circ}$ using the angle scans.
For all structure analyses, the program suite SHELX was used for structure solution and least-squares refinement \cite{Sheldrick}.
Platon was used to check for missing symmetry elements in the structures \cite{Spek}.
The obtained parameters are shown in Tables~I and II.
It was fairly consistent with those of the previous polycrystalline samples \cite{Rundqvist2,Liu2}.
The occupancies estimated at each site were more than 99\%, ensuring the high quality of the crystal.

\begin{table}[htb]
\caption{Crystallographic data of single-crystal Mn$_3$P at room temperature. }
\begin{tabular}{|c|c|}\hline
Formula & Mn$_3$P \\
Crystal system & tetragonal \\
Space group & $I$\=4 (no.82) \\
$a$ (\AA) & 9.182 \\
$c$ (\AA) & 4.5655 \\
$Z$ & 8 \\
Unique reflections & 579 \\
Residual factor $R1$ & 0.0299 \\
$wR2$ & 0.0679 \\ \cline{1-2}
\end{tabular}
\end{table}

\begin{table}[htb]
\caption{Structural parameters of single-crystal Mn$_3$P at room temperature.  The l.s. indicates local symmetry.}
\begin{tabular}{|cccccccc|}\hline
site & wyckoff & l.s. & x & y & z & $B_{eq}$ (\AA$^2$) & occup. \\ \cline{1-8}
Mn1 & 8g & 1 & 0.58059 & 0.60739 & 0.7714 & 0.49 & 1 \\
Mn2 & 8g & 1 & 0.85655 & 0.53203 & 0.0137 & 0.39 & 0.99 \\
Mn3 & 8g & 1 & 0.67197 & 0.71927 & 0.2468 & 0.43 & 1 \\
P & 8g & 1 & 0.7937 & 0.54463 & 0.5109 & 0.43 & 1 \\ \cline{1-8}
\end{tabular}
\end{table}

\section{Experimental methods}

Magnetic susceptibility measurement was performed in a range of $2-300$ K by utilizing a Magnetic Property Measurement System (Quantum Design).
Specific heat measurement was performed in a range of $2-50$ K by a Physical Property Measurement System (Quantum Design).
We measured electrical resistivity using four probe AC method, in which electrical contacts of the wire were made by a spod-weld method.
The high pressure up to 3.5 GPa for resistivity measurements was applied using an indenter-type pressure cell and Daphne 7474 oil as a pressure-transmitting medium \cite{indenter,Murata}.
The low temperature for resistivity measurements was achieved using a homemade $^3$He cryostat, except for the data at 1.38 and 1.63 GPa, where a homemade dilution refrigerator was used.
In the $^3$He cryostat, the pressure cell was soaked in the liquid, while it was cooled by thermal conduction in the dilution refrigerator.
The temperature was measuremd from the resistivity of a calibrated RuO$_2$ thermometer.

The neutron diffraction measurements at ambient and high pressures were performed on the time-of-flight diffractometer CORELLI at the Spallation Neutron Source at Oak Ridge National Laboratory.\cite{Ye}
High pressure up to 1 GPa was generated with a self-clamped piston-cylinder cell made of CuBe utilizing Daphne 7373 oil. The crystal dimensions were about 1$\times$1$\times$3 mm$^{3}$. 
The pressure was monitored by measuring the lattice constant of a comounted NaCl crystal.
The neutron diffraction measurements at ambient pressure were also performed on a triple-axis spectrometer HB-1 at the High Flux Isotope Reactor at Oak Ridge National Laboratory. Neutrons with an energy of 13.5 meV were used, together with a horizontal collimator sequence of 48’-80’-S-80’-240’. Contamination from higher-order beams was effectively eliminated using pyrolytic graphite filters. The single crystal was oriented in the (H, H, L) scattering plane and mounted in a closed-cycle $^4$He gas refrigerator.

We also performed NMR measurements through a conventional spin-echo method and band calculations through a full-potential LAPW (linear augmented plane wave) calculation within the LDA(local density approximation).
The results of NMR and the band calculations are shown in the following sections in the Supplemental Material. 

\section{Nuclear magnetic resonance for powdered sample}

\setcounter{figure}{0}

\begin{figure}[htb]
\centering
\includegraphics[width=0.8\linewidth]{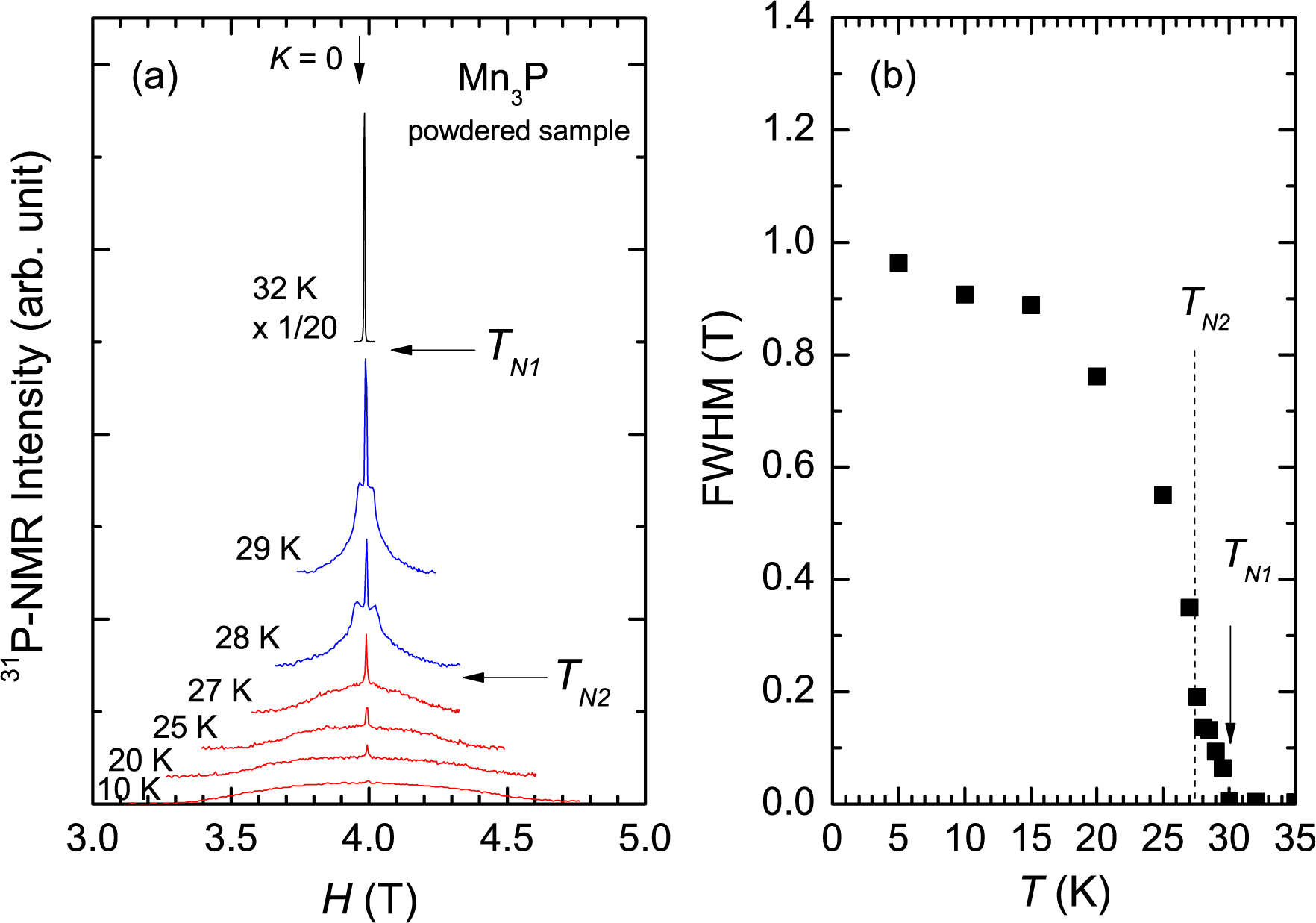}
\caption[]{(a) $^{31}$P-NMR spectra for powdered single crystals of Mn$_3$P. The measurements were done with fixing frequency at $f=68.35$ MHz. The symmetric broadening with the center at $K=0$ position evidences the AF ordered state. The unexplained sharp signal with small volume fraction remains even below $T_{N2}$, indicating that it arises from some impurity phase. (b) Temperature dependence of the internal field at the P site. The continuous and successive development are observed through $T_{N1}$ and $T_{N2}$.}
\end{figure}

We performed NMR measurements to check that the double magnetic transition is intrinsic. 
Figure 1(a) shows $^{31}$P-NMR spectra measured at several temperatures using the unoriented powdered crystals.
In the paramagnetic (PM) state, the sharp spectrum is observed near the position of zero Knight shift, $K$.
Below $T_{N1}=30$ K, the spectrum is significantly broadened, and the symmetric broadening with respect to the $K=0$ position indicates the AF arrangement of the ordered moments.
The sharp signal, where the internal field is quite weak, remains even below $T_{N1}$ and also $T_{N2}$.
A volume fraction of the remaining signal is about 5$\%$ at 29 K, which is too small to recognize that it is intrinsic. 
The intensity is gradually weakened as decreasing temperature, and the sharp signal does not change its shape when crossing $T_{N2}$.
The origin of this remaining signal is unclear, but it is not relevant with the second transition at $T_{N2}$.
What happened at $T_{N2}$ is a change in the shape of the spectrum, which is already broadened below $T_{N1}$.
Figure 1(b) displays the temperature dependence of the full-width at the half maximum, which is estimated with omitting the remaining sharp signals. 
Obviously, a two-step change is observed in the spectral width, which is consistent with the intensity of magnetic scattering shown in Fig.~3(a) in the main paper, suggesting that the both second-order like transitions occur at the identical P sites as the intrinsic property.

\section{Band structure calculation}

Band structure calculations were obtained through a full-potential LAPW (linear augmented plane wave) calculation within the LDA(local density approximation).
Calculated density of states (DOS) for the PM state is shown in Fig.~2, where the partial DOS of the respective Mn sites and the P site are shown.
The DOS at the Fermi energy originates in the $3d$ orbitals of three Mn sites, and the contributions from each Mn site are similar.
The $3d$-DOS including all $3d$ orbitals roughly forms the rectangle shape with a bandwidth of $\sim0.3$ Ry $\sim$ 4 eV, indicative of the itinerant character of the $3d$ electrons.
$\gamma_{\rm band} = 16.3$ mJ/mol K$^2$ is estimated in the PM state for a formula cell.

\begin{figure}[htb]
\centering
\includegraphics[width=0.9\linewidth]{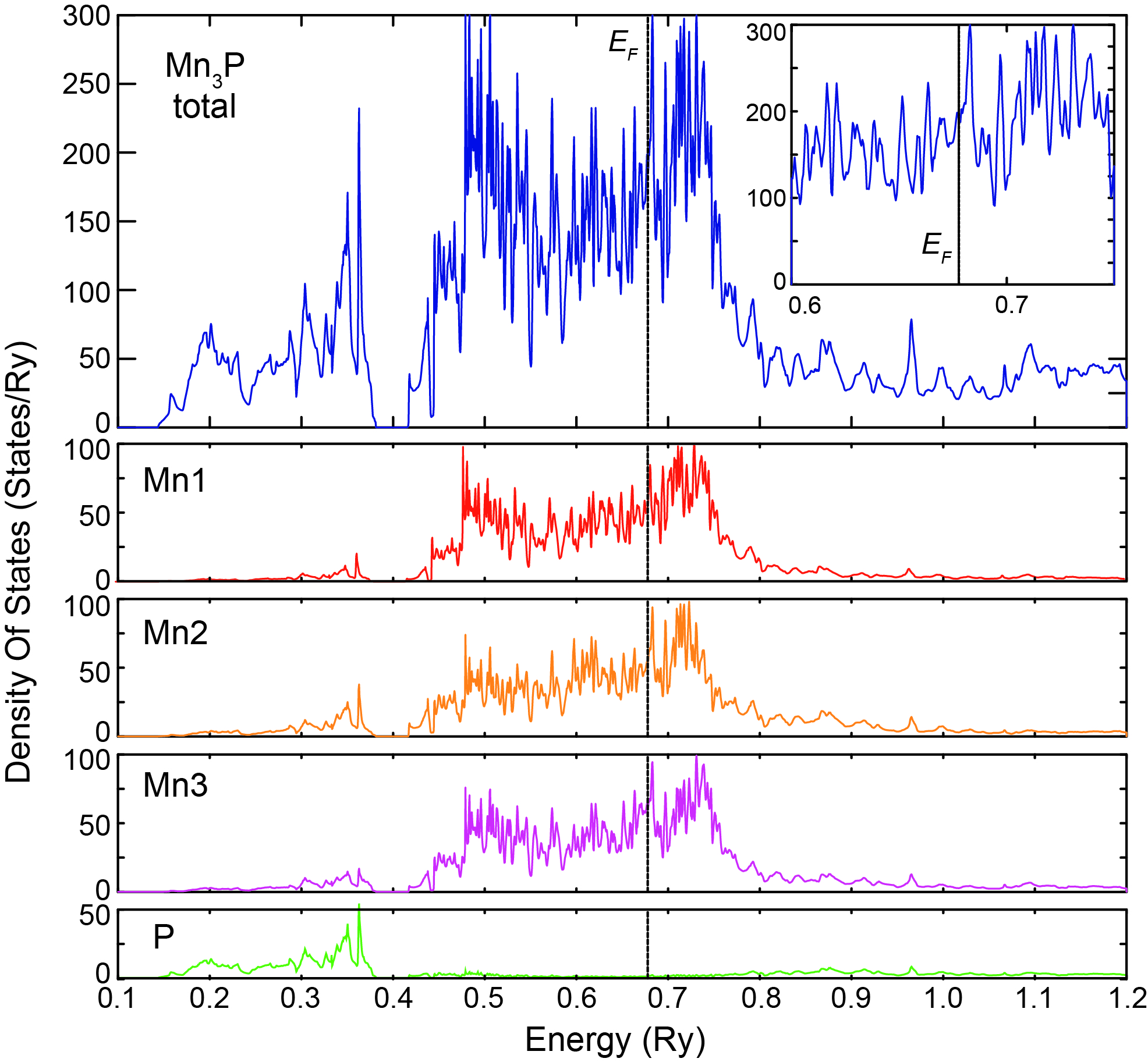}
\caption[]{Density of states (DOS) of Mn$_3$P calculated for the PM state. The partial DOS from the respective Mn sites and the P site are shown by different colors. The $4s$ contribution of Mn and $3s$ contribution of P are omitted, because they are negligibly small. The DOS at the Fermi energy is composed of $3d$ electrons of three Mn sites. The rectangle shape of the $3d$-DOS reflects the itinerant character of Mn$_3$P. The $\gamma_{\rm band} = 16.3$ mJ/mol K$^2$ is estimated.}
\end{figure}

\section{Crystal and magnetic-structure analyses}

The neutron diffraction measurements were performed on a time-of-flight neutron diffractometer CORELLI installed at Spallation Neutron Source at Oak Ridge National Laboratory. 
The single crystal was mounted with ($HK$0) in the horizontal scattering plane in the pressure cell. Thanks to the wide coverage of detectors vertically, Bragg reflections with finite $L$ were observed.
The crystal and magnetic-structure analyses were performed using the FullProf \cite{fullprof2} and Jana \cite{Jana2} packages.

\begin{figure}[htb]
\centering
\includegraphics[width=0.9\linewidth]{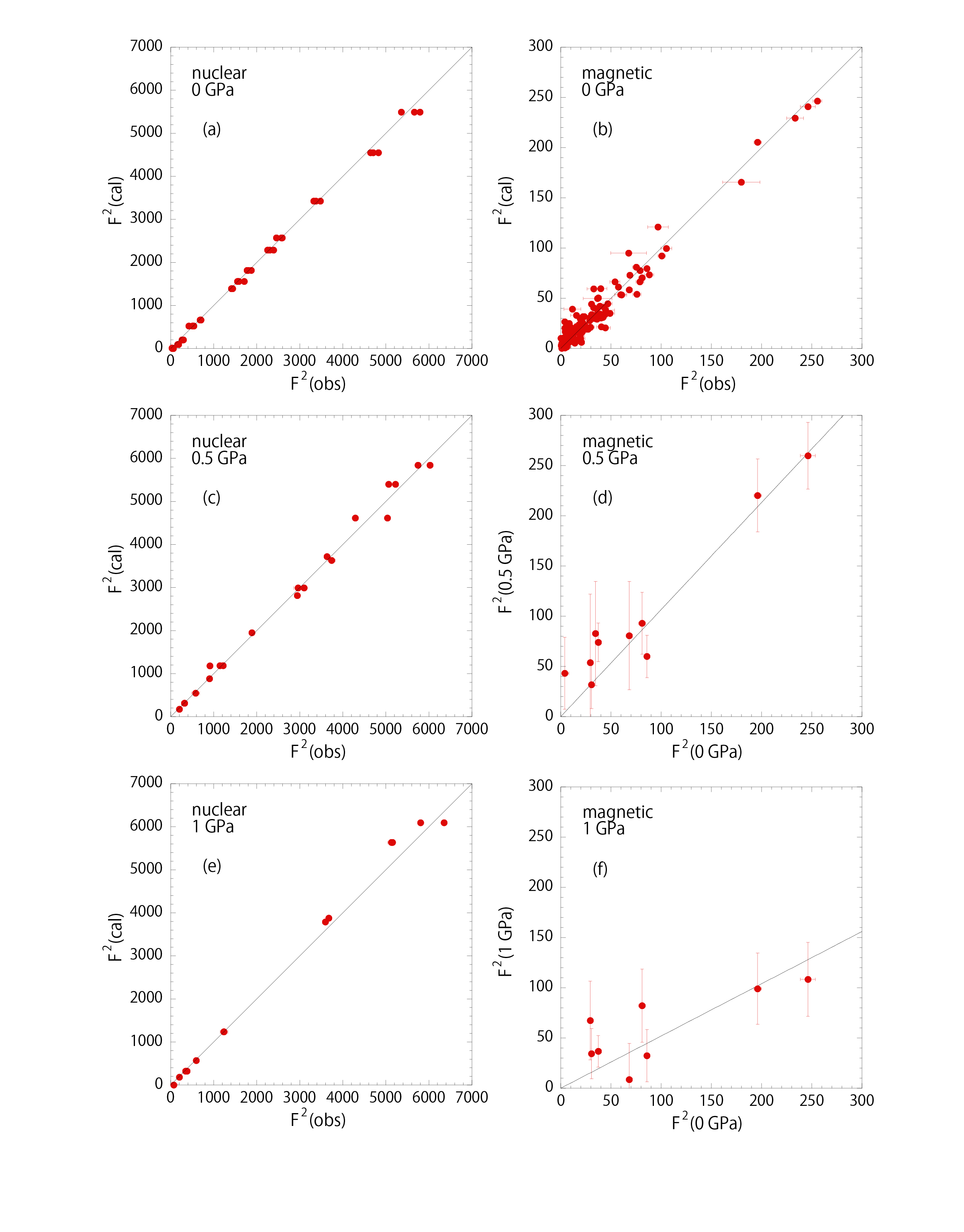}
\caption[]{(a)-(c) and (e) are plots of $F\rm_{calc}^2$ vs. $F\rm_{obs}^2$ measured at 7 K. (a), (c), and (e) display nuclear Bragg reflections at 0 ($R_{F}$=3.68\%), 0.5 ($R_{F}$=2.40\%), and 1 GPa ($R_{F}$=4.24\%), respectively. (b) displays magnetic Bragg reflections at 0 GPa ($R_{F}$=11.3\%). The solid lines correspond to $F^2\rm_{cal}$=$F^2\rm_{obs}$.
(d) and (f) show the relation between $F\rm_{obs}^2$ at ambient and high pressures. The solid lines are the results of fits to a linear curve.}
\end{figure}

The observed Bragg reflection intensity ($I_{\rm obs}$) is proportional to the squared structure factor ($F\rm_{obs}^2$).
\begin{equation}
I_{\rm obs}= C\cdot N(\lambda)\cdot A(\lambda)\cdot Y(\lambda,\theta,F{\rm_{cal}})\cdot L(\lambda,\theta)\cdot f(\lambda,\theta)^2\cdot F\rm_{obs}^2,
\label{intensity}
\end{equation}
where $C$, $N$, $\lambda$, $A$, $Y$, $L$, $\theta$, $F{\rm_{cal}}$, and $f$ are constant, incident neutron beam flux, wavelength, sample absorption, extinction correction, Lorentz factor, half the scattering angle, calculated structure factor, and magnetic form factor, respectively. $f$ is unit for nuclear Bragg peaks.
We first performed crystal structure analysis at each pressure to determine $C$ and $Y$ in Eq. (1). As shown in Figs. 3(a), (c), and (e), the analysis of the nuclear Bragg intensities was performed reasonably well. 
The detailed magnetic structure analysis was performed at ambient pressure. 
The magnetic structure models were evaluated using the representation analysis with the program BasIreps in Fullprof Suite \cite{suite} to determine the symmetry-allowed magnetic structure. 
The magnetic structure, shown in Figs. 1 and 4 in the main paper, was determined. 
The fitting was reasonably good, as shown in Fig. 3(b).

At 0.5 and 1 GPa, the number of observable magnetic Bragg peaks is quite limited due to large background signal and low beam transmission originating from the CuBe pressure cell. Therefore, we did not perform magnetic structure analysis for the data at the two high pressures. Instead, the observed magnetic Bragg intensities were compared with those at ambient pressure and the overall reduction factor for the magnetic moments was obtained, assuming that the overall magnetic structure does not change. Figs. 3(d) and (f) show the relation of observed magnetic structure factors at ambient and high pressures. The linearity is reasonably good at 0.5 and 1 GPa, suggesting that the above assumption is adequate. The square root of the slope of the linear curve corresponds to the reduction factor of the overall magnetic moments. The factor is 1.03(10) and 0.72(7) at 0.5 and 1 GPa, respectively. The magnetic moments do not change at 0.5 GPa but slightly decrease at 1.0 GPa.

\section{Change in the wave vector at $T_{N2}$}

Figure 4 shows the magnetic Bragg intensities just below $T_{N2}=27.5$ K. The wave vectors estimated from these data are plotted in Fig.~3(b) in the main paper. The wave vector decreases below $T_{N2}$ without any signature of broadening or a splitting of the peak.  The experimental data suggets that the transition at $T_{N2}$ is continuous within the experimental resolution.

\begin{figure}[htb]
\centering
\includegraphics[width=0.9\linewidth]{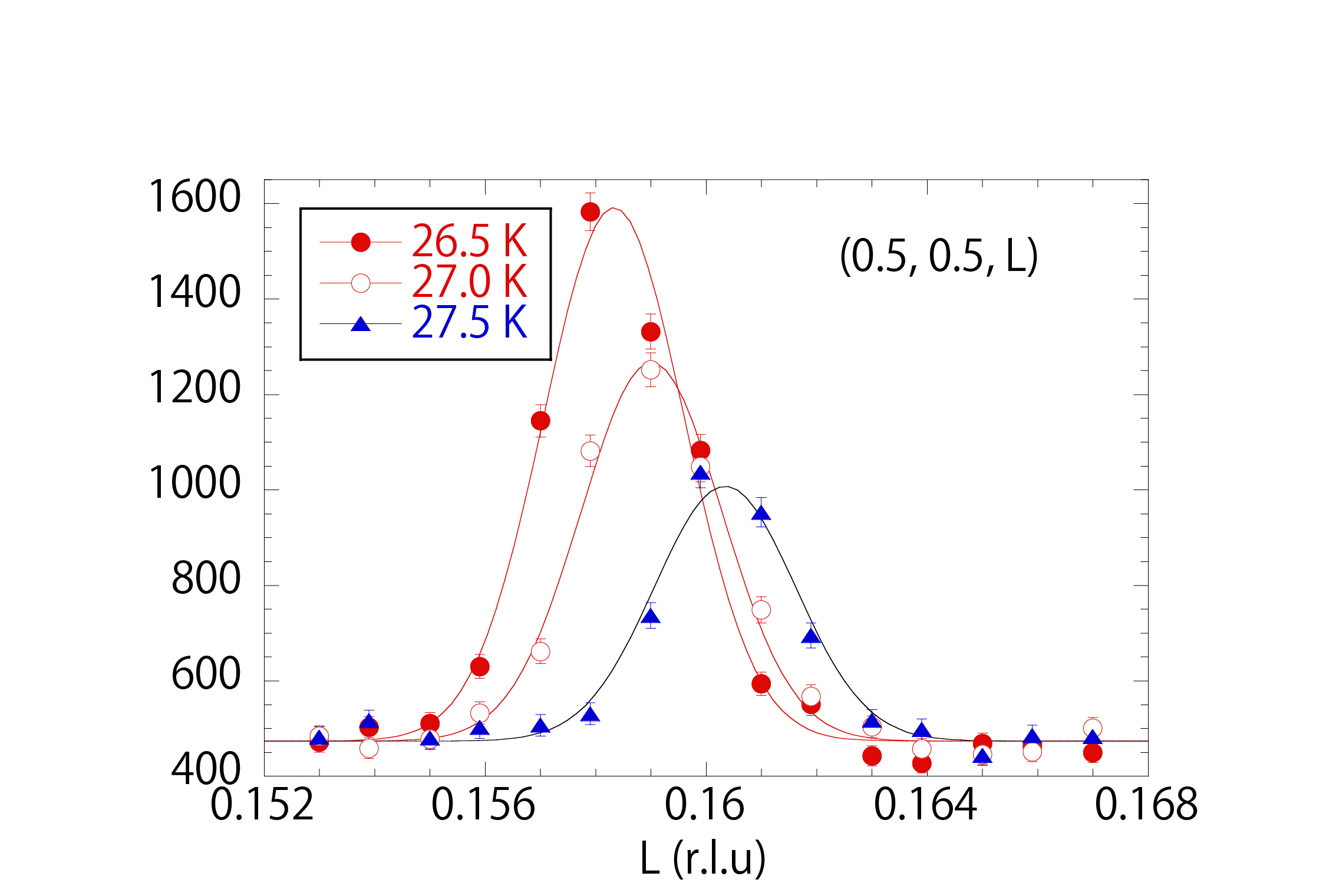}
\caption[]{Magnetic Bragg intensities at (0.5, 0.5, L) for several temperatures. The change in the wave vector below $T_{N2}=27.5$ K is shown.}
\end{figure}

\section{Magnetic structure in the intermediate phase}

\begin{figure}[htb]
\centering
\includegraphics[width=0.7\linewidth]{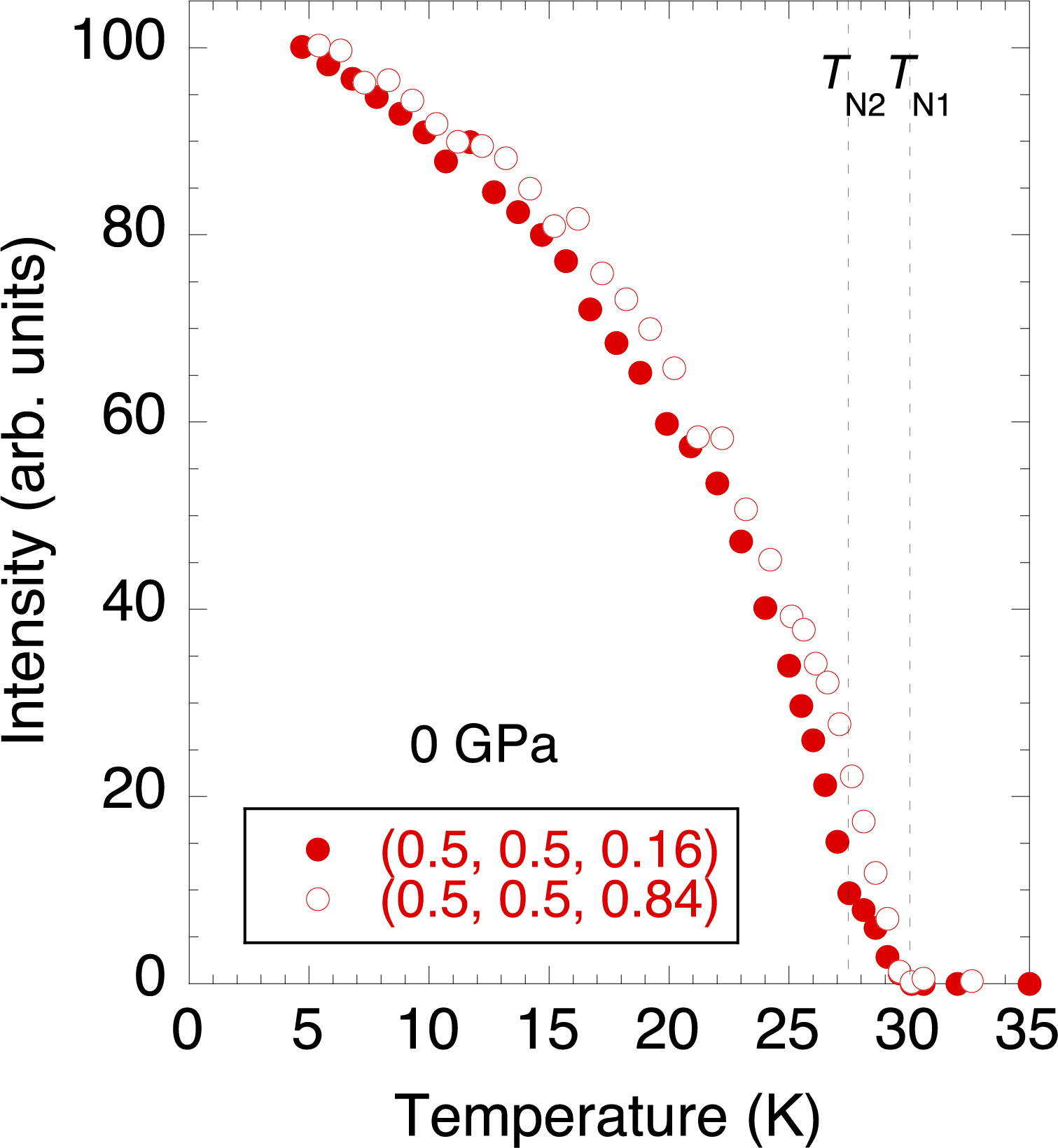}
\caption[]{Temperature dependence of integrated intensities of the magnetic Bragg peaks around (0.5, 0.5, 0.16) and (0.5, 0.5, 0.84) measured at ambient pressure. The (0.5, 0.5, 0.16) and (0.5, 0.5, 0.84) intensities are normalized at 5 K.}
\end{figure}

As described in the main paper, the magnetic structure analysis in the intermediate phase is challenging, since the magnetic Bragg intensities are very weak. In order to understand the magnetic structure in the intermediate phase qualitatively, we measured detailed temperature dependence of the two magnetic Bragg intensities around (0.5, 0.5, 0.16) and (0.5, 0.5, 0.84). The (0.5, 0.5, 0.16) reflection consists of 83\%\ from the spin component in the $ab$ plane ($M_{ab}$) and 17\%\ from the spin component along the $c$ axis ($M_c$). On the other hand, the (0.5, 0.5, 0.84) reflection consists of 15\%\ from $M_{ab}$ and 85\%\ from $M_c$.  Therefore, the former and latter reflections represent mostly $M_{ab}^2$ and $M_c^2$, respectively. [The (0.5, 2.5, 0.16) reflection, shown in Fig. 3(a) of the main paper, consists of 98.4\%\ from $M_{ab}$ and 1.6\%\ from $M_c$.] Figure 5 shows temperature dependence of magnetic Bragg peaks around (0.5, 0.5, 0.16) and (0.5, 0.5, 0.84) normalized at 5 K. We found that the (0.5, 0.5, 0.84) intensities are more than two times stronger than the (0.5, 0.5, 0.16) intensities in the intermediate phase, suggesting that $M_c$ is more dominant in the intermediate phase. If the Mn2b site also carries large $M_c$ in this phase as in the low temperature phase, the Mn2b moments might order mostly below $T_{N1}$ and other moments might order below $T_{N2}$.

\section{Systematic resistivity data under pressure}

Figure 6 shows the $\rho$ vs $T^2$ plot for various pressures. 
At ambient pressure, the $\rho$ obeys a $T^2$ dependence at low temperatures, but it is acheived only below $\sim1.5$ K.
The estimated $A$ coefficient is $0.52 \mu\Omega$cm/K$^2$.
The temperature range of the $T^2$ dependence shrinks near the QCP ($P_c\sim1.6$ GPa).
The large $A$ is maintained up to the pressure just above the QCP, and the $A$ decreases at higher pressure.

\begin{figure}[htb]
\centering
\includegraphics[width=0.8\linewidth]{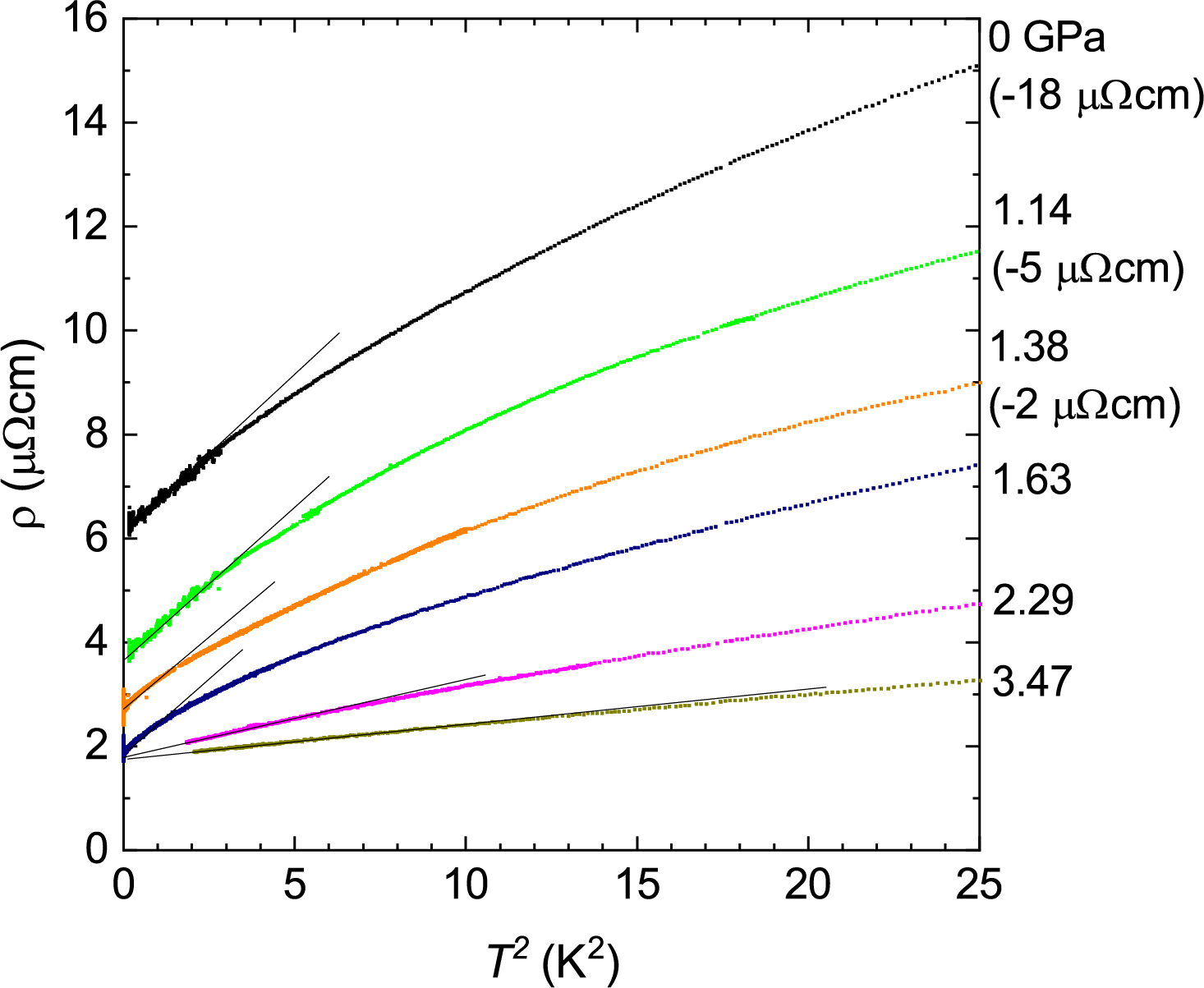}
\caption[]{$\rho$ vs $T^2$ plot for various pressures. The data at lower pressures are shifted, because $\rho_0$ is large in the ordered state. The straight line is used for the estimationof the $A$ coefficient.}
\end{figure}

\end{document}